%% file: main.tex
\documentclass[envcountsame]{llncs}

\usepackage[english]{babel}
\usepackage[T1]{fontenc}
\usepackage{lmodern}
\usepackage[utf8]{inputenc}
\usepackage[final]{microtype}

\usepackage[l2tabu, orthodox]{nag}
\usepackage[ruled,noline]{algorithm2e}

\usepackage{bbm}
\usepackage{proof}
\usepackage{csquotes}
\usepackage{mathtools}
\usepackage{booktabs}
\usepackage{multirow}
\usepackage{afterpage}

\usepackage{tikz}
\usetikzlibrary{automata,positioning,shapes,calc,fit,arrows.meta,backgrounds}
\usepackage{pgfplots}
\usepgfplotslibrary{fillbetween}
\pgfplotsset{compat=1.13}

\tikzstyle{state}+=[minimum size=20pt,inner sep=2pt]
\tikzstyle{action}=[font=\small,inner sep=0pt,outer sep=3pt]
\tikzstyle{actionedge}=[->,draw]
\tikzset{chainarrow/.tip={Stealth[length=3pt]}}
\tikzset{>=chainarrow}

\tikzstyle{system} = [draw,circle,minimum size=1cm]
\tikzstyle{environment} = [draw,diamond,minimum size=1cm]

\tikzstyle{systemltl} = [draw,rectangle,rounded corners=5pt,minimum size=1cm]
\tikzstyle{environmentltl} = [draw,rectangle,minimum size=1cm]

\tikzset{every picture/.append style={
	auto,
	node distance=2cm,
	initial text={},
}}

\pgfplotscreateplotcyclelist{no marks}{%
	loosely dotted, mark=\empty\\
	densely dotted, mark=\empty\\
	solid, mark=\empty\\
	densely dashed, mark=\empty\\
	loosely dashed, mark=\empty\\
	dashdotted, mark=\empty\\
	dashdotdotted, mark=\empty\\
	densely dashdotted, mark=\empty\\%
}

\usepackage[shortcuts]{extdash}
\usepackage{enumitem}

\input{0_macros}
\usepackage[capitalise]{cleveref}
\usepackage{subfig}

\advance\textwidth2mm 
\advance\hoffset-1mm

\newcommand{\citex}[1]{}

\title{Semantic Labelling and Learning for\\ Parity Game Solving in LTL Synthesis\thanks{This research was funded in part by the Czech Science Foundation grant No.~\mbox{P202/12/G061}, and the German Research Foundation (DFG) projects KR 4890/1-1 \enquote{Verified Model Checkers} and KR 4890/2-1 \enquote{Statistical Unbounded Verification}.}}
\author{
	Jan K\v{r}et{\'i}nsk{\'y} \orcidID{0000-0002-8122-2881}
	\and Alexander Manta \orcidID{0000-0003-0591-6985}
	\and Tobias Meggendorfer \orcidID{0000-0002-1712-2165}
}

\institute{Technical University of Munich}

\begin{document}

\pagestyle{plain}
\maketitle
\begin{abstract}
\input{0_abstract}
\end{abstract}
\input{1_introduction}
\input{2_preliminaries}
\input{3_ideas}
\input{5_evaluation}
\input{8_conlusion}

\bibliographystyle{plain} 
\bibliography{main}
\newpage
\appendix
\input{9_appendix}

\end{document}

%% file: 0_abstract.tex

We propose ``semantic labelling'' as a novel ingredient for solving games in the context of LTL synthesis.
It exploits recent advances in the automata-based approach, yielding more information for each state of the generated parity game than the game graph can capture.
We utilize this extra information to improve standard approaches as follows.
(i)~Compared to strategy improvement (SI) with random initial strategy, a more informed initialization often yields a winning strategy directly without any computation.
(ii)~This initialization makes SI also yield smaller solutions.
(iii)~While Q-learning on the game graph turns out not too efficient, Q-learning with the semantic information becomes competitive to SI.
Since already the simplest heuristics achieve significant improvements the experimental results demonstrate the utility of semantic labelling.
This extra information opens the door to more advanced learning approaches both for initialization and improvement of strategies.

%% file: 1_introduction.tex
\section{Introduction}

\newcommand{\para}[1]{\paragraph{#1}}

\para{Reactive synthesis} is a classical problem\citex{Church,DBLP:conf/icalp/PnueliR89,RamadgeWonham87} to find a strategy that given a stream of inputs gradually produces a stream of outputs so that a given specification over the inputs and outputs is satisfied.
In LTL synthesis\citex{DBLP:conf/icalp/PnueliR89} the specification is given as a formula of \emph{linear temporal logic (LTL)}\citex{Pnueli77}.
The classical solution technique is the \emph{automata-theoretic approach} \citex{Buchi62}\cite{DBLP:conf/lics/VardiW86} that transforms the specification into an automaton. 
The partitioning of atomic propositions into inputs and outputs then yields a game over this automaton. 
Subsequently,  the game is solved and the winning strategy in the game induces a winning strategy for the original problem.
The standard type of automaton to be used in this context is the \emph{deterministic parity automaton (DPA)} since (i) determinism ensures we obtain a well-defined game and (ii) the parity condition yields a \emph{parity game}, which can be solved reasonably cheaply in practice \cite{DBLP:journals/tcs/Zielonka98,DBLP:conf/fsttcs/Schewe07,DBLP:conf/cav/Fearnley17}\citex{,DBLP:conf/stoc/CaludeJKL017,DBLP:conf/spin/FearnleyJS0W17} with good tool support \cite{DBLP:conf/atva/FriedmannL09,DBLP:conf/tacas/Dijk18}.

While solving large games still takes significant resources, the bottleneck of this procedure is already the construction of the game.
Indeed, after transforming the LTL formula into a non-deterministic automaton \cite{DBLP:conf/lics/VardiW86} this automaton is determinized using \emph{Safra's construction}  \cite{DBLP:conf/focs/Safra88} or its improvements \cite{DBLP:conf/lics/Piterman06,DBLP:conf/fossacs/Schewe09}.
This is notoriously known to be practically inefficient \cite{DBLP:conf/sofsem/Kupferman12} despite some tool support \cite{DBLP:conf/wia/KleinB07}\citex{,ltl2dstar}. 
As a result, other approaches for the synthesis problem have been suggested that avoid Safra's determinization, see the related work below.
However, recent translators, e.g. \cite{DBLP:conf/atva/Duret-LutzLFMRX16,DBLP:conf/cav/KretinskyMSZ18}, yield significantly smaller automata to the extent that the traditional parity-game approach, e.g. \cite{michaud.18.synt,DBLP:conf/cav/MeyerSL18}, becomes competitive \cite{DBLP:journals/corr/abs-1711-11439,Syntcomp18}.
Apart from the smaller sizes of automata, one of their decisive advancements is the ability to generate the automaton on the fly and terminate as soon as a winning strategy is found, possibly way earlier than the whole automaton is constructed.

Yet these approaches suffer from several inefficiencies.
In order to tackle them let us observe their roots.
Firstly, despite the relative efficiency, solving the parity game can still take significant time.
Either the whole game is solved, e.g. using Zielonka's algorithm \cite{DBLP:journals/tcs/Zielonka98} as in \cite{michaud.18.synt}, or growing on-the-fly explored parts are repetitively solved, e.g. using strategy improvement as in \cite{DBLP:conf/cav/MeyerSL18}.
In both cases, large parts of the state space are processed and the overall effort is still significant since the strategy improvement is executed many times during the process.
Secondly, in the case with on-the-fly exploration, it is not clear in which direction the game should be explored.
In the graph game, the available extra information is only the priorities and computing their attractors is a global computation defeating the purpose of on-the-fly exploration.

In this paper, we suggest a framework for a theoretically fundamental improvement of solving parity games that arise from LTL synthesis and we instantiate it with the first simple heuristics.
The experimental results confirm the potential of this approach.
The main idea is to exploit, to our best knowledge for the first time, the \emph{semantics} of the vertices of the game.

Where does the semantics come from?
Since the original specification is translated to an automaton, its states have a strong correspondence to the monitored property.
However, Safra's determinization and the subsequent latest appearance record for obtaining a DPA leaves us with permutation over Safra's trees over sets of LTL formulae, whose semantics is extremely hard to decipher.
In contrast, the new approach of \cite{DBLP:conf/cav/EsparzaK14,
	DBLP:conf/cav/KretinskyMSZ18} yields a description of each state in terms of a single formula to be satisfied and a list of formulae describing progress of satisfying each sub-goal.
This clearer structure allows us to exploit the meaning of available successors and to choose the most promising one in the sense of satisfying the goal of each player.
This addresses issue of exploration guidance.
The other issue of updating the whole or whole explored part of the state space can be addressed using reinforcement learning \cite{RL}.
Since the degree how promising a vertex is can be quantified, we can use it as a reward, together with the priorities, in Q-learning.
This way we update only the most promising parts of the state space.

\para{Our contribution} is the following:
\begin{itemize}[noitemsep,topsep=0pt,parsep=0pt,partopsep=0pt]
	\item We introduce a semantics-based framework for heuristics for parity games in LTL synthesis and instantiate it as follows.
	\item We utilize the semantic labelling of vertices to get a better initial strategy for the parity game, often yielding (i) an optimal solution directly and (ii) a smaller one.
	\item We show how reinforcement learning can be applied to parity games and accelerated by the semantic labelling.
	\item We demonstrate the potential of this approach experimentally on formulae from the SYNTCOMP competition \cite{DBLP:journals/corr/abs-1711-11439} as well as random formulae, opening a door for learning-based approaches to automata-based LTL synthesis.
\end{itemize}

%
%
%
%
%
%
%
%
%
%
%
%
%
%

\subsection*{Related Work}

Firstly, one can reduce the synthesis problem to emptiness of nondeterministic B\"uchi tree automata~\cite{DBLP:conf/focs/KupfermanV05};
it has been implemented with considerable success in~\cite{DBLP:conf/fmcad/JobstmannB06}.
The second approach is to use heuristic to improve Safra's determinization
procedure~\cite{DBLP:journals/tcs/KleinB06,DBLP:conf/wia/KleinB07}. 
The third approach is to consider fragments of LTL. 
For instance, the generalized reactivity(1) fragment of LTL (called GR(1))  was introduced 
in~\cite{DBLP:conf/vmcai/PitermanPS06} and a cubic time symbolic representation of an equivalent automaton was 
presented. The approach has been implemented in the ANZU tool~\cite{DBLP:conf/cav/JobstmannGWB07}.
Another approach, prominent in competitions like SYNTCOMP \cite{DBLP:journals/corr/abs-1711-11439}, is bounded synthesis \cite{DBLP:conf/atva/ScheweF07a}, as implemented by, e.g., BoSy \cite{DBLP:conf/cav/FaymonvilleFT17} and PARTY \cite{DBLP:conf/cav/KhalimovJB13}.
The tool Acacia+ \cite{DBLP:conf/cav/BohyBFJR12} uses symbolic incremental algorithms based on antichains.

Besides, there are learning approaches that utilize some information on the state space.
However, this is typically not the information on the property currently to be satisfied, e.g. \cite{DBLP:conf/tacas/NeiderT16} uses automata learning, but only for safety properties and not focusing on the property itself.
Further, \cite{DBLP:journals/automatica/DingLB14} takes the property into account, but only as its respective automaton. 
It tries to decrease the distance to the accepting vertex, which can be very different from making the property easier to satisfy.
Moreover, it is not designed for games, although the alternating distance might address this drawback.
More importantly, it is not suited for partial models as we need to construct the whole automaton, which is the bottleneck for complex properties.

%% file: 2_preliminaries.tex
\section{Preliminaries}

In this section, we give some basic background knowledge and establish fundamental notation.
Due to space constraints, we touch only briefly on several topics and encourage the reader to refer to the mentioned literature.

\paragraph{Basic Notation.}
We use $\Naturals$ to denote the set of non-negative integers.
Given a propositional formula $\phi$ over a set of propositions $\AP$, we use $\satisfying(\phi) = \{v \in 2^{\AP} \mid v \models \phi\}$ to denote the set of all satisfying assignments.
The constants $\true$ and $\false$ denote \emph{true} and \emph{false}, respectively.

\subsection{Synthesis \& Games}

The synthesis problem in its general form asks whether a system can be controlled in a way such that it satisfies a given specification under any (possible) environment.
Moreover, one often is interested in obtaining a witness to this query, i.e. some \emph{controller} or \emph{strategy} which specifies the system's actions.
For example, one might ask whether a robot can be steered over difficult terrain such that it arrives at a particular target location.

\begin{figure}[t]
	\centering
	\begin{tikzpicture}[auto,node distance=2cm,initial text=]
		\node[systemltl,initial left] (s0) {$v_0, 4$};
		\node[environmentltl] [right of=s0] (s1) {$v_1, 2$};
		\node[systemltl] [below of=s1] (s2) {$v_2, 1$};
		\node[environmentltl] [right of=s1] (s3) {$v_3, 3$};
		\node[systemltl] [below of=s3] (s4) {$v_4, 5$};

		\path[->]
			(s0) edge[loop above] (s0)
			(s0) edge (s1)
			(s0) edge (s2)

			(s1) edge (s3)
			(s1) edge[bend right] (s2)

			(s2) edge[bend right] (s1)
			(s2) edge (s3)

			(s3) edge[loop above] (s3)
			(s3) edge (s4)

			(s4) edge[loop right] (s4)
		;
	\end{tikzpicture}

	\caption{An example (parity) game.
	Rounded rectangles belong to the system player and normal rectangles to the environment player.
	The vertices are additionally labelled with their priorities.
	For readability, we omit the requirement of alternation.} \label{fig:example_game}
\end{figure}

\paragraph{Graph games} are a standard formalism used in synthesis.
A game, denoted by $\GraphGame = ((\Vertices, \Edges), \initialvertex, \VertexPlayer, \WinningCondition)$, consists of a digraph $(\Vertices, \Edges)$, a \emph{starting vertex} $\initialvertex \in \Vertices$, a \emph{player mapping} $\VertexPlayer$, and a \emph{winning condition} $\WinningCondition$, described later.
Each vertex belongs to one of the two players $0$ (called \emph{system}) and $1$ (called \emph{environment}), specified by the mapping $\VertexPlayer : \Vertices \to \{0, 1\}$.
In other words, the set of vertices is partitioned into player $0$'s vertices $\VerticesSys$ and player $1$'s vertices $\VerticesEnv$; $\Vertices = \VerticesSys \strictunion \VerticesEnv$.
See \cref{fig:example_game} for an example of such a game.
For ease of notation, we write $v \Edges := \{(v, u) \in \Edges \mid u \in \Vertices\}$ to denote all outgoing edges of some vertex $v$ and define $E_i := \{(u, v) \in \Edges \mid u \in \Vertices_i\}$ the set of all edges \enquote{controlled} by player $i$.

To play the game, a token is placed in the initial vertex $\initialvertex$.
Then, the player owning the token's current vertex moves the token along an outgoing edge of the current vertex.
This is repeated infinitely, giving rise to an infinite sequence of vertices containing the token $\play = \initialvertex v_1 v_2 \cdots \in \Vertices^\omega$, called a \emph{play}.
The set of all possible plays is denoted by $\Plays$.

For simplicity, we assume in the following that all games are \emph{alternating}, i.e. the successors of a vertex belong to a different player than the vertex itself.

\paragraph{Winning conditions} are a mapping from plays to the winning player $\WinningCondition : \Plays \to \{0, 1\}$.
Numerous kinds of winning conditions have been studied.
In this work, we consider the following three:
\begin{description}[noitemsep,topsep=0pt,parsep=0pt,partopsep=0pt]
	\item[Safety] is defined by a set of vertices $T$ to be avoided.
		The system player loses iff one of the vertices in the given set is visited.
	\item[Co-Safety] (or \emph{reachability}) is, as the name suggests, dual to safety.
		Here, the system player wins iff one of the given vertices is visited at least once.
		Observe that this exactly corresponds to a safety objective for the environment player.
	\item[Parity] is based on a \emph{priority assignment} for each vertex $\priorityAssignment : \Vertices \to \Naturals$.
		The system player wins iff the \emph{maximum} of all infinitely often occurring priorities is \emph{odd}.\footnote{Instead of the maximum, one could also decide based on the minimum; similarly instead of \enquote{odd}, \enquote{even} sometimes is considered winning for the system.}
		Formally, we define $\inf(\play) = \{p \mid \forall j. \exists k \geq j. \priorityAssignment(\play_k) = p\}$ and system wins a play $\play$ iff $\max \inf(\play)$ is odd.
		We refer to odd priorities as \emph{good} (for the system player) and to even priorities as \emph{bad} (for the system player).
\end{description}
Note that both safety and co-safety are special cases of the parity condition, with a straightforward linear time transformation.

\paragraph{Strategies} are mappings $\strategy_i : V_i \to E$, assigning to each of the player's vertices an edge along which the token will be moved.\footnote{Strategies may be significantly more complex, e.g., by using memory.
Since \enquote{positional} strategies are sufficient for all properties we consider, we intentionally omit the general definition in the interest of space.}
Observe that once both players fix a strategy, the game is fully determined and a unique run is induced.
This means that given a game with a particular winning condition and a strategy for each player, we can decide which of the players wins the game using these strategies.
A strategy of a player is called \emph{winning} if the player wins for \emph{any} strategy of the opponent.
Thus, we can rephrase the synthesis question to \enquote{Is there a winning strategy for the system player?}. 

For example, consider again the game depicted in \cref{fig:example_game}.
Fixing the strategies $\strategy_0 = \{v_0 \mapsto (v_0, v_2), v_2 \mapsto (v_2, v_3), v_4 \mapsto (v_4, v_4)\}$ and $\strategy_1 = \{v_1 \mapsto (v_1, v_2), v_3 \mapsto (v_3, v_3) \}$ induces the play $v_0 v_2 v_3 v_3 \cdots$.
The set of infinitely often seen priorities equals $\{3\}$, hence the system player wins with these strategies.
Moreover, the strategy $\strategy_0$ is winning, since the play always ends up in either $v_3$ or $v_4$.

\paragraph{Strategy Improvement} (or \emph{strategy iteration}) is the most prominent way of solving parity games, i.e. answering the above question.
In recent times, it received significant attention due to both theoretical\citex{DBLP:conf/spin/FearnleyJS0W17,DBLP:conf/stoc/CaludeJKL017} and practical advances\citex{DBLP:conf/atva/HoffmannL13,DBLP:conf/cav/Fearnley17,DBLP:conf/atva/MeyerL16,DBLP:conf/atva/FriedmannL09}.
We explain the approach only very briefly, since its details are not important for this work.

In essence, strategy improvement works as follows.
First, arbitrary initial strategies are picked for both players.
Then, the algorithm checks whether one of the current strategies is winning.
If yes, this strategy is returned.
Otherwise, the algorithm tries to improve one of the strategies by changing its choices in some vertices.
If an improvement is not possible, there exists no winning strategy for the respective player.
Otherwise, the process is repeated with the new strategy.

It is known that this algorithm converges to the correct result in finite time for any initial strategy.
This gives us a straightforward way of optimization, namely the choice of the initial strategy.
Intuitively, a heuristic which often comes up with a \enquote{good} strategy may improve the runtime significantly over arbitrary or random choice, since then only a few improvement steps are necessary.

Throughout this work, we refer to a reference implementation of SI, denoted \texttt{SI}, e.g., when running SI with a particular initial strategy.
In our implementation, we used the algorithm of \cite{DBLP:conf/cav/VogeJ00}, but other variants could be substituted.

\subsection{Linear Temporal Logic}

\emph{Linear Temporal Logic} (LTL) is a standard logics used in verification and synthesis to specify the desired behaviour of a system.
The logic is given by the syntax
\begin{equation*}
	\phi ::= \lfalse \mid a \mid \lnot \phi \mid \phi \land \phi \mid \ltlNext \phi \mid \phi \ltlUntil \phi,
\end{equation*}
where $a \in \AP$ is an \emph{atomic proposition}, inducing the \emph{alphabet} $\Alphabet = 2^\AP$.
LTL formulae are interpreted over infinite sequences $w \in \Alphabet^\omega$ called $\omega$-words.
Intuitively, a word $w = w_0 w_1 \cdots \in \Alphabet^\omega$ satisfies the \emph{next} $\ltlNext \phi$ iff $\phi$ is satisfied in the next step.
The \emph{until} operator $\phi \ltlUntil \psi$ is satisfied iff $\phi$ holds until $\psi$ is satisfied.
Apart from the mentioned operators, we also consider \emph{finally} $\ltlFinally \phi := \ltrue \ltlUntil \phi$ and \emph{globally} $\ltlGlobally \phi := \lnot \ltlFinally \lnot \phi$, which require that $\phi$ holds at least once or always, respectively.

Given an LTL formula $\phi$, the set of its \emph{sub-formulae} is denoted by $\subformulas(\phi)$.
The \emph{top-level temporal operators} $\topformulas(\phi)$ are all temporal operators not nested inside other temporal operators.
For example, the formula $\phi = \ltlGlobally ((\ltlFinally a) \land b) \land \ltlFinally b$ has sub-formulae $\subformulas(\phi) = \{a, b, \ltlFinally a, (\ltlFinally a) \land b, \ltlGlobally ((\ltlFinally a) \land b), \ltlFinally b, \phi\}$ and top-level operators $\topformulas(\phi) = \{\ltlGlobally ((\ltlFinally a) \land b), \ltlFinally b\}$.

\paragraph{LTL Synthesis} is an instance of the general synthesis problem.
Here, the specification to be satisfied by the system is given in form of an LTL formula.
Due to recent advances \citex{DBLP:conf/atva/GaiserKE12,DBLP:conf/atva/KretinskyL13,DBLP:conf/atva/KomarkovaK14,}\cite{DBLP:conf/cav/KretinskyMSZ18,DBLP:conf/cav/MeyerSL18}, the \emph{automata-based approach} \cite{DBLP:conf/lics/VardiW86} to LTL synthesis received significant attention.
Essentially, the given LTL formula is translated into an $\omega$-automaton, which in turn is transformed into a parity game.
By solving the resulting game, we obtain a solution to the original synthesis question.

Technically, the game is obtained by \enquote{splitting}.
To this end, the set of atomic propositions is split into system- and environment-controlled propositions.
Then, the players' actions correspond to choosing which of their propositions to enable.
Once both players chose their propositions' values, the automaton moves to the next vertex according to the chosen valuation.
See, e.g., \cite{DBLP:conf/cav/MeyerSL18}, for more detail.

\paragraph{Semantic translations} from LTL to automata are the key ingredient to our new approach.
These translations not only produce a parity game, but also provide a semantic labelling of the game's vertices.
In particular, using the 
approach introduced in \cite{DBLP:conf/cav/EsparzaK14}
and implemented in \cite{DBLP:conf/cav/KretinskyMSZ18}, we obtain for each vertex a list of LTL formulae, roughly corresponding to (sub-)goals which still have to be (possibly repetitively) fulfilled.
Due to space constraints, we describe the ideas of these constructions only briefly in \cref{sec:contributions:input}.
This labelling is not easily derived from the structure of the game graph and provides additional information not accessible to conventional, general-purpose parity game solvers.
The primary goal of this paper is to show that this additional information can be exploited for a significant increase in performance.

\subsection{Q-Learning}

Q-Learning is a well known, simple yet versatile \emph{reinforcement learning} technique \cite{RL}.
It usually is applied in machine learning to find performant strategies for (stochastic) systems.
The technique roughly works as follows.
Each edge $(u, v)$ has a \emph{Q-value} $Q(u, v)$, indicating the \emph{quality} assigned to the respective edge.
This value is initialized according to some heuristic and then repeatedly updated through \emph{learning episodes}.
Each episode consists of sampling a path through the system, following the maximal Q-value.
In order to encourage exploration, randomization is added to this choice.
While sampling, the learning agent receives rewards based on his choices.
The Q-value of the respective edge is then updated with the obtained reward and the Q-value of it's successor.\footnote{The exact details of this update vary between different instantiations of Q-learning.
For example, a discount factor may be included.}
To smoothen the learning process, the propagated value is weighted by a \emph{learning rate} $\alpha$.
Together, the update essentially is computed by $Q(v, u) \gets (1 - \alpha) \cdot Q(v, u) + \alpha \cdot \left( \mathcal{R}(v, u) + Q(u) \right)$ where $v$ and $u$ are the current and next vertex, respectively, $\mathcal{R}(v, u)$ is the obtained reward, and $Q(u) = \max_{(u, u') \in E} Q(u, u')$ is the Q-value of the successor vertex $u$.


%% file: 3_ideas.tex
\section{Our Contributions}

In this section, we explain the central ideas of our contributions.
First, we highlight the peculiarities of the mentioned labelling function.
Then, we introduce the concept of \emph{trueness}, which we directly use to augment strategy improvement.
Finally, we explain our adaptation of Q-learning to parity games and how we derive a semantic reward signal from the labelling, using trueness.

\subsection{Input Details} \label{sec:contributions:input}

We assume that we are given a parity game, where each vertex is labelled by a structured list of LTL formulae.
The labelling corresponds to the remaining goals to be achieved by the system player (or violated by the environment player).
More precisely, the labelling consists of one \emph{master formula} and potentially several \emph{monitors}.
The master formula tracks the \enquote{overall progress} and all finitely achievable parts of the formula.
In particular, the master formula \true{} corresponds to the formula being satisfied by the prefix, analogously \false{} corresponds to a falsified formula.
The system player automatically wins if a vertex labelled with \true{} is reached and, similarly, loses on a \false{} vertex.
In the special case of reachability or safety specifications, the labelling actually only consists of the master formula.

For more complex specifications, the labelling also exhibits a more intricate structure.
Intuitively, there is one monitor for each sub-formula which needs to be satisfied infinitely often (liveness conditions) or may only be violated finitely often (safety conditions).
Each monitor tracks a list of formulae which have to be fulfilled in order to satisfy its overall goal.
The monitors are ordered according to an appearance-record style construction.
\enquote{Failing} monitors emit a bad priority and are moved at the beginning of the list, succeeding monitors instead emit a good priority.
Both priorities are based on the respective monitor's position in the list.
Intuitively, for a fixed word $\omega$, all monitors which only fail finitely often eventually are ranked higher than all the monitors which fail infinitely often.
Thus, if such a non-failing monitor emits a good priority, it overrules all of the failing monitors' bad priorities.

Consider, for example, the formula $\ltlFinally \ltlGlobally a$, meaning \enquote{eventually, $a$ appears every step}.
Here, there is no ultimate \true{} or \false{} vertex, since this formula cannot be satisfied or violated by any finite prefix.
The construction gives us $\ltlFinally \ltlGlobally a$ as master formula and $\ltlGlobally a$ as a monitored goal.
Whenever we see a $\lnot a$, this monitor would fail, emit a bad priority, and move to the front of the list.
Dually, for every $a$, it emits a good priority.

The details of this construction are described in \cite{DBLP:conf/tacas/EsparzaKRS17}.
It is implemented in the tool \texttt{Rabinizer} \cite{DBLP:conf/cav/KretinskyMSZ18} 
which we use for our constructions.
An online demo thereof is located at \url{https://owl.model.in.tum.de/try/}. 
A simplified example can be found in \cref{fig:example_labelling} later on.

\subsection{(Co-)Safety games}

Recall that for these games, the labelling we obtain is a single LTL formula per vertex.
The system player wants to reach the $\true$ vertex and avoid the $\false$ vertex.
Consequently, the system player naturally is interested in taking \enquote{trueness-maximizing} edges, analogously the environment player wants to move away from $\true$.
This simple observation directly leads us to the concept of \emph{trueness}.

\subsubsection{Trueness} of an LTL formula $\trueness : \LTL \to [0, 1]$ intuitively denotes how \enquote{close} a given formula is to being satisfied.
We compute this value by treating the formula as purely propositional, i.e. each temporal operator is considered to be a fresh variable.
Formally, given an LTL formula $\phi$, we interpret it as a propositional formula over $\Alphabet = 2^{\AP \union \topformulas(\phi)}$.
For this formula, we then determine and scale the ratio of satisfying assignments, i.e.\ $\trueness(\phi) := \cardinality{\satisfying(\phi)} / \cardinality{\Alphabet}$.
This value can be computed efficiently by representing the formula as \emph{binary decision diagram (BDD)}\citex{DBLP:reference/mc/Bryant18}, as implemented in \texttt{Rabinizer}.
Even for formulae with several hundred syntax elements the trueness is computed virtually instantaneously.

At first, this notion may seem rather unintuitive, since the temporal aspect of a formula is not necessarily reflected by the trueness value.
For example, the formula $\phi = \ltlGlobally a \land \ltlGlobally \lnot a$ has a trueness value of $\trueness(\phi) = \frac{1}{4}$, but actually is unsatisfiable.
Nevertheless, trueness proves to be a surprisingly good initialization heuristic, which we explain in the following and further demonstrate in our evaluation.

One particular reason for its performance in our application is due to the way the labelling is constructed.
In particular, temporal operators are \enquote{unfolded} as an essential step of the construction.
For example, the formula $\ltlGlobally a$ is unfolded to $a \land \ltlGlobally a$ while $\ltlFinally a$ yields $a \lor \ltlFinally a$ and $a \ltlUntil b$ gives $b \lor (a \land (a \ltlUntil b))$.
The unfolded variants are semantically equivalent to the original formula, but provide us with a one-step propositional \enquote{approximation} of the temporal operators.
The above $\phi$ then is unfolded to $\phi \equiv (a \land \ltlGlobally a) \land (\lnot a \land \ltlGlobally \lnot a) \equiv \false$.

\subsubsection{Initializing strategies based on trueness} can be achieved as follows.
Fix some game $\GraphGame$ with a (co-)safety objective and labelling function $L : \Vertices \to \LTL$.
The strategies $\strategy_0$ and $\strategy_1$ are called \emph{trueness-optimal} if they satisfy
\begin{align*}
	\strategy_0(v_0) & \in \argmax_{(v_0, u) \in \Edges} \trueness(L(u)) & \strategy_1(v_1) & \in \argmin_{(v_1, u) \in \Edges} \trueness(L(u))
\end{align*}
for all vertices $v_0 \in \VerticesSys$ and $v_1 \in \VerticesEnv$, respectively.
Observe that, since we assumed the game to be alternating, all $u$ vertices in the above equations belong to the respective opponent.

This immediately yields our first semantic algorithm $\texttt{SI}_{sem}$, which runs \texttt{SI} initialized with trueness optimal strategies.

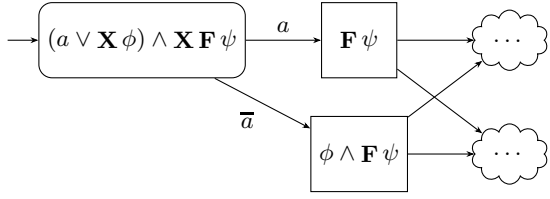
\begin{figure}[t]
	\centering
	\begin{tikzpicture}[auto,node distance=2cm,initial text=]
		\node[systemltl,initial left] (s0) {$(a \lor \ltlNext \phi) \land \ltlNext \ltlFinally \psi$};
		\node[environmentltl] [right=1cm of s0] (s1) {$\ltlFinally \psi$};
		\node[environmentltl] [below=0.5 cm of s1] (s2) {$\phi \land \ltlFinally \psi$};
		\node[cloud, draw, cloud puffs=10,aspect=1.5] [right of=s1] (s1dots) {$\dots$};
		\node[cloud, draw, cloud puffs=10,aspect=1.5] [right of=s2] (s2dots) {$\dots$};

		\path[->]
			(s0) edge node {$a$} (s1)
			(s0) edge[swap] node {$\overline{a}$} (s2)

			(s1) edge (s1dots)
			(s1) edge (s2dots)
			(s2) edge (s1dots)
			(s2) edge (s2dots)
		;
	\end{tikzpicture}

	\caption{
		An example showing the application of trueness initialization.
		For readability, we only show the vertex labels.
		$\phi$ and $\psi$ are some non-trivial LTL formulae.
	} \label{fig:example_safety}
\end{figure}

For a small example, consider the simplified part of a game depicted in \cref{fig:example_safety}.
Here, the system player can choose whether or not to play $a$ in the initial vertex.
The successors' trueness value, namely $\trueness(\ltlFinally \psi) = \frac{1}{2}$ and $\trueness(\phi \land \ltlFinally \psi) \leq \frac{1}{4}$ (for a non-trivial $\phi$), suggest the natural choice of $a$, leading to $\ltlFinally \psi$.
Intuitively, $\ltlFinally \psi$ is \enquote{easier} to satisfy than $\phi \land \ltlFinally \psi$.
Observe that without the labelling and its trueness value this choice would not be as obvious, since the impact of this decision may only become visible much later in the game.
Quite surprisingly, this initialization solves a \emph{majority} of randomly generated games \emph{instantly} without the need for any further improvements step, as shown in the experimental evaluation.

%

\subsection{Parity games}

To apply our ideas to parity games, there are several hurdles to overcome.
Recall that the labelling we obtain for these games has a non-trivial structure, compared to the singleton labelling in the special cases.
Hence, we cannot use the trueness value directly.
Rather, we need a more intricate way of deriving meaning from the labelling.
Because of its simplicity, we decided to use Q-learning.
Recall that Q\=/learning usually works with a single agent, interested in maximizing the obtained reward.
In our case, we instead have two antagonistic agents and we need to adapt the Q-learning framework to this setting.
Furthermore, there are some technical peculiarities when we want to incorporate priorities and the labelling.

\begin{remark}
	We again stress that our main goal is to show the usefulness of the so far unexploited semantic labelling.
	In particular, the exact approach to extracting rewards or even the fact that we use Q-learning is of secondary importance for our research goal.
	Using more advanced techniques, e.g., extract reward from the formula using neural networks, is left for future work.
\end{remark}

\subsubsection{Basic concepts}

For all variants, our Q-values lie between $-1$, corresponding to a presumably guaranteed loss for the system player, and $+1$, analogously corresponding to a win.
As usual in Q-learning, we repeatedly sample paths and update vertex values.
But, since the two players are antagonistic, we don't always pick a maximizing action while sampling.
Instead, we choose a successor with maximal Q-value in system vertices and a minimal one in environment vertices.
Once we encounter a vertex for a second time and consequently would enter a loop, we stop the sampling.

In the following we present three variants of Q-learning.
The first, agnostic variant obtains rewards only based on whether an episode is winning or losing.
This approach is widely applicable, since basically no domain knowledge is necessary.
Not surprisingly, it also is not too efficient.
We then present a first adaption, which additionally incorporates priorities.
This variant is tailored towards parity games, but does not employ any semantic information provided by the labelling.
Our experiments show that it outperforms the first variant, but only by a slim margin.
Finally, we present our semantic approach, which on top also considers the labelling, employing the trueness function in several ways.
In our experiments, this significantly outperforms the first two ideas and, on some datasets, even beats strategy improvement by a large factor.
Note that each variant also incorporates the reward signals of the previous approaches.


\subsubsection{Rewards based on winning paths} are binary, yielding $+1$ for winning and $-1$ for losing.
In our case, this means that when we stop sampling after entering a loop, we determine whether the loop is winning or losing and propagate the value accordingly.
We initialize the Q-values with $0$, since there is no a-priori information available.
The resulting Q-learning variant is denoted by $\texttt{QL}_{win}$.

\subsubsection{Priority rewards} are the first step to a more intricate reward signal.
Here, the agent additionally obtains intermediate rewards based on the priority of the edge.
Recall that these priorities are natural numbers, and the system player is interested in large, odd priorities.
Dually, large even priorities should be avoided.

There are two difficulties associated with this idea.
Since we assumed our Q\=/values to lie between $-1$ and $+1$, we need to rescale the priorities into this domain.
Furthermore, we need to rescale them such that larger priorities significantly \enquote{overrule} smaller ones, reflecting the nature of the parity objective.
For example, obtaining ten $5$ priorities in a row is irrelevant if afterwards one $6$ is encountered.

We approach this problem by rescaling the priorities as follows.
Let $p_i$ be the priorities occurring in the given game, sorted in ascending order.
We first compute the absolute frequency of each priority as $f(p_i) = \cardinality{\{s \mid \priorityAssignment(s) = p_i\}}$, i.e. the number of vertices in the whole game labelled with priority $p_i$.
Then, we define the scaled priorities $\overline{p}_i$ by $\overline{p}_0 = p_0$, $\overline{p}_i = 2 \cdot f(p_{i-1}) \cdot \overline{p}_{i-1} + 1$.
Observe that $\overline{p}_i$ is larger than the sum of all $\overline{p}_j$ with $j < i$ occurring in the game.
Finally, we re-normalize $\overline{p}_i$ into the $[-1, +1]$ domain by $r_i := (-1)^{p_i + 1} \overline{p}_i (1 + \sum_j \overline{p}_j \cdot f(p_j))^{-1}$
Even priorities are mapped to the negative domain, as the system player wants to avoid them.
As before, we initialize the Q-values with $0$.

We denote this Q-learning variant by $\texttt{QL}_{pri}$.
Note that this approach is applicable to general parity games.
In our case, it indirectly uses the labelling, since the priorities of the game are directly derived from the labelling and correspond to progress in the monitors.

\subsubsection{Semantic rewards} are our idea for exploiting the information provided by the vertex labelling, denoted by $\texttt{QL}_{sem}$.
Firstly, we describe how we assign the initial Q-values.
Recall that we cannot apply the trueness value to the whole labelling, since in general it comprises several different LTL formulae corresponding to different goals.
Nevertheless, we can still easily exploit the master formula to obtain a sensible value.
In particular, we initialize the Q-value of each action based on the trueness of the master formula in the successor vertex.
This directly generalizes the approach of the special case of (co-)safety, where the labelling consists only of the master formula.

\begin{figure}[t]
	\centering
	\begin{tikzpicture}
		\node[systemltl,initial left] (s0) {$\ltlGlobally \ltlFinally (a \land \ltlNext b); [a \land \ltlNext b]$};
		\node[systemltl,right=1.5cm of s0] (s1) {$\ltlGlobally \ltlFinally (a \land \ltlNext b); [b]$};

		\path[->]
			(s0) edge[out=20,in=160] (s1)
			(s0) edge[loop above] (s0)

			(s1) edge[out=200,in=-20] (s0)
			(s1) edge[loop above] (s1)
		;
	\end{tikzpicture}
	\caption{A simplified example for a monitor labelling as produced by \texttt{Rabinizer}.
	The first formula represents the master formula, while the second one corresponds to the only monitor's formula.
	} \label{fig:example_labelling}
\end{figure}
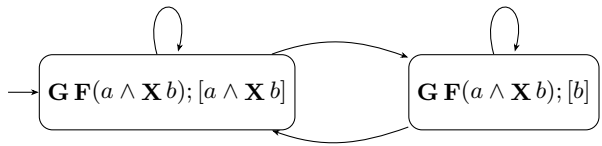

Now, we introduce our ideas for deriving the reward signal from the labelling.
Recall that apart from the master formula we have several monitor formulae, corresponding to goals which we have to fulfil repeatedly.
We present a motivating example in \cref{fig:example_labelling}.
There, we show a (simplified) labelling produced for the formula $\psi = \ltlGlobally \ltlFinally (a \land \ltlNext b)$.
In order to satisfy $\psi$, we repeatedly need to play $a$ and then in the next step $b$.
The master formula is the same in both vertices, hence we need to analyse the monitors instead.
Intuitively, playing $a$ seems to be the more natural choice in the initial vertex, since $b$ is easier to satisfy than $a \land \ltlNext b$.
Hence, our main idea is to also apply trueness analysis to the monitor labels.


Consequently, the system player should be interested not only in the ad-hoc reward obtained by following good priorities, but also in \enquote{progressing} other monitors.
To this end, we also give a reward proportional to the change in trueness for all monitors which are ranked higher than the current priority as follows.

Let $e = (u, v)$ be an edge in some game $\GraphGame$ with priority assignment $\priorityAssignment$ and labelling $L$.
By taking this edge, each monitor updates its monitored formulae.
In particular, some monitor may fail or succeed due to this transition.
Let $m_e$ be the index of the highest-ranked of such failing or succeeding monitors in the labelling of vertex $u$.
If there is no such monitor, let $m_e = -1$.
Then, all monitors ranked higher than $m_e$ made some progress, but did not fail or succeed.
Nevertheless, the system player is interested in succeeding at least one of those monitors and let none of them fail.
Hence, we incentivise the system player to take actions which additionally bring monitors closer to succeeding or at least do not worsen their state.
As basis for this approximation, we again use the trueness value.
Note that we still use the winning and priority based rewards, hence the \enquote{progress} of the monitor at $m_e$ is already incorporated.

Each monitor is a list of formulae which all have to be fulfilled repeatedly.
Since we want all of these goals to be satisfied, we first take the minimum trueness value for all formulae tracked by a particular monitor.
Furthermore, since the progress of all monitors ranked lower than $m_e$ is irrelevant due to the success or fail event of $m_e$, we ignore them.
Together, we are left with one value for each monitor ranked higher than $m_e$.
To aggregate these values, we pick the maximal one, letting the learner focus on improving a single monitor instead of slowly progressing all of them.
The overall \enquote{progress} reward thus is given by
\begin{equation*}
	\mathsf{progress} = \max_{m > m_e} \left( \min_{\phi \in L(v)_m} \trueness(\phi) - \min_{\phi \in L(u)_m} \trueness(\phi) \right),
\end{equation*}
where $L(v)_m$ are the formulae tracked by monitor $m$ in vertex $v$.

Note that this is only one of many possible choices.
Since we are only interested in showing the general applicability of our idea, we simply picked the best heuristic out of several hand-crafted definitions of $\mathsf{progress}$.

%% file: 5_evaluation.tex
\section{Experimental Evaluation}

In this section, we evaluate the presented techniques to show their potential.
We show how initializing strategy improvement using semantic information leads to significant improvements of the algorithm.
We evaluate our Q-learning variants on several data sets, both real-world and randomly generated.
We compare it to strategy improvement, showing our semantic variant to be competitive on several models.
Further data, left out due to space constraints, can be found in \cref{sec:appendix:data}.

\subsection{Setup}

The experiments have been carried out on consumer-grade hardware, a laptop with a 2x2,9 GHz Intel Core i5 and 8GB RAM.
We investigate several algorithms and models, which we briefly explain in the following.

\subsubsection{Algorithms}

In our evaluation, we investigate the following algorithms:
\begin{itemize}[noitemsep,topsep=0pt,parsep=0pt,partopsep=0pt]
	\item $\texttt{SI}$: A reference SI implementation with random initial strategy.
	\item $\texttt{SI}_{sem}$: SI with semantic initialization.
	\item $\QL_{win}$: Q-learning with only win/loss as reward signal.
	\item $\QL_{pri}$: Q-learning with priority-based rewards.
	\item $\QL_{sem}$: Q-learning with semantic rewards.
\end{itemize}
While running the Q-learning variants, we repeatedly check whether the current strategy is winning in the starting vertex in order to determine when to stop the learning.
We do not use any information gained by this check during the learning process itself.

\subsubsection{Metrics}
First, we count the \emph{number of evaluation steps} until convergence for each algorithm.
Since our implementation is only a prototype, we consider time to be less relevant, and use this metric instead to approximate the time needed by an efficient implementation of each variant.
For Q-learning, this equates to the number of vertices visited in all learning episodes; for SI we count the number of iterations times the size of the game, to allow for a fair comparison, giving a slight advantage to SI to be on the safe side.
See the below remark for further details.

Second, we investigate the \emph{size of the solution}, i.e.\ the number of vertices reachable under the identified winning strategy.
The size of a solution is a good estimate for its quality.
For example, a smaller solution means that its implementation requires less memory, since decisions need to be stored for fewer states.
We give the solution size as a fraction of the overall size of the respective game.

As both methods involve randomization -- Q-learning during sampling, SI in its initialization -- we ran our methods five times on each model.
We chose a timeout of 60 seconds for each run to allow for a reasonably fast evaluation.
Since timeouts are difficult to properly incorporate into averages, we chose to ignore the few timeouts that occurred, usually less than $5\%$.
See \cref{sec:appendix:data} for details.

\begin{remark}
	Q-learning and SI evaluate the strategy in vastly different ways.
	Q\=/learning picks the currently best action whenever it visits a particular vertex, potentially switching back and forth between two similar actions.
	In contrast, strategy improvement repeatedly evaluates the current strategy on the whole game, simultaneously changing choices in all vertices which allow for improvement.
	Intuitively, Q-learning evaluates fewer, important vertices more often, while the evaluations of SI are spread over the whole game.
	The evaluation of a strategy in SI is costly, since the whole game is considered, while Q-learning simply compares and updates the current Q-values along a single path.
\end{remark}

\subsubsection{Models}

We investigate both randomly generated and real-world games.

The random formulae are generated using Spot's \cite{DBLP:conf/atva/Duret-LutzLFMRX16} \texttt{randltl}.
We selected three classes of random formulae:
\begin{itemize}[noitemsep,topsep=0pt,parsep=0pt,partopsep=0pt]
	\item (Co-)Safety: Pure safety or co-safety formulae
	\item Near(Co-)Safety: Formulae which mostly consist of either safety or co-safety elements, but contain a few sub-formulae of the other type.
	\item Parity: Fully random formulae.
\end{itemize}
We investigate the special case of \enquote{Near(Co-)Safety} formulae separately, since a lot of real-world specifications often comprise mostly safety and only a few other conditions.
This dataset is supposed to imitate this asymmetry.

In order to generate these formulae, we parametrize \texttt{randltl} with priorities on the syntax elements.
For the \enquote{Parity} dataset, we used the default priorities.
For the other two, we used the priorities listed in \cref{sec:appendix:data}.
Furthermore, to obtain a reasonable test-set, we filter the generated formulae as follows.
First, we remove all formulae where the translation to a parity game using \texttt{Rabinizer} takes more than 5 GB of memory or more than 30 seconds, as this would lead to disproportionately large games.
Then, we also remove games which have more than $10{,}000$ nodes.
We generated 100 such formulae per class.

We also use several real-world formulae from the SYNTCOMP 2017 competition \cite{DBLP:journals/corr/abs-1711-11439}.
The specifications are given in the \emph{TLSF}\citex{DBLP:journals/corr/Jacobs016} format, which \texttt{Rabinizer} can translate to LTL and then to parity games.
Again, we filter out games with more than $10{,}000$ nodes, leading to a total of 195 models.

\subsection{Results}

\begin{table}[!t]
	\setlength{\tabcolsep}{5pt}
	\caption{
		Percentage of games solved in the starting vertex by the initial strategy and the size of the final solutions.
		\enquote{Near} refers to the Near(Co-)Safety dataset.
		To obtain more deterministic results, we used additional semantic information obtained from the monitors for tie-breaking in $\texttt{SI}_{sem}$, where applicable.
	} \label{tbl:results_simple}
	\centering
	\begin{tabular}{l|ccc||ccc}
		                    & \multicolumn{3}{c||}{Immediately solved games} & \multicolumn{3}{c}{Solution size} \\
		                    & (Co-)Safety & Near &          Parity           & (Co-)Safety & Near &    Parity    \\
		\hline
		$\texttt{SI}$       &    32\%     & 11\% &           10\%            &     7\%     & 13\% &     8\%      \\
		$\texttt{SI}_{sem}$ &    65\%     & 67\% &           56\%            &     7\%     & 13\% &     9\%
	\end{tabular}
\end{table} %
\begin{figure}[!t]
	\centering

	\vspace{1em}
	\begin{tikzpicture}
		\begin{axis}[width=\textwidth,height=2cm,
				xlabel=\empty,ylabel=\empty,
				legend pos={outer north east},
				cycle list name=no marks,
				hide axis,
				xmin=0,xmax=0,ymin=0,ymax=0,
				legend style={
					at={(current bounding box.center)},
					anchor=center,
					legend columns=-1
				},
				legend entries={
					{$\QL_{win}$},
					{$\QL_{pri}$},
					{$\QL_{sem}$},
					{$\texttt{SI}$},
					{$\texttt{SI}_{sem}$},
				},
				cycle list name=no marks,
			]
			\pgfplotsinvokeforeach{1,...,5}{\addplot coordinates {(0,0)};}
		\end{axis}
	\end{tikzpicture}
	\vspace{-1em}

	\subfloat[(Co-)Safety]{
		\begin{tikzpicture}
			\begin{axis}[xmin=0,xmax=2100,ymax=1.0,samples=50,width=0.4\linewidth,height=5cm,
					axis x line = middle, axis y line* = middle,
					enlarge y limits=0,
					xlabel=\empty,ylabel=\empty,
					xtick={0,500,1000,1500,2000},xticklabels={,500,,1500},
					ytick={0,0.2,0.4,0.6,0.8,1.0},yticklabels={0,0.2,0.4,0.6,0.8,1.0},
					legend pos={outer north east},
					cycle list name=no marks
				]
				\addplot+ plot table[col sep=comma] {data/random-reach-safe-ql.txt};
				\addplot+ plot table[col sep=comma] {data/random-reach-safe-ql_prio.txt};
				\addplot+ plot table[col sep=comma] {data/random-reach-safe-ql_sem.txt};
				\addplot+ plot table[col sep=comma] {data/random-reach-safe-si.txt};
				\addplot+ plot table[col sep=comma] {data/random-reach-safe-si_sem.txt};
			\end{axis}
		\end{tikzpicture}
	}
	\subfloat[Near(Co-)Safety]{
		\begin{tikzpicture}
			\begin{axis}[xmin=0,xmax=2100,ymax=1.0,samples=50,width=0.39\linewidth,height=5cm,
					axis x line = middle, axis y line* = middle,
					enlarge y limits=0,
					xlabel=\empty,ylabel=\empty,
					xtick={0,500,1000,1500,2000},xticklabels={,500,,1500},
					ytick={0,0.2,0.4,0.6,0.8,1.0},yticklabels={},
					legend pos={outer north east},
					cycle list name=no marks
				]
				\addplot+ plot table[col sep=comma] {data/pseudo-reach-safe-ql.txt};
				\addplot+ plot table[col sep=comma] {data/pseudo-reach-safe-ql_prio.txt};
				\addplot+ plot table[col sep=comma] {data/pseudo-reach-safe-ql_sem.txt};
				\addplot+ plot table[col sep=comma] {data/pseudo-reach-safe-si.txt};
				\addplot+ plot table[col sep=comma] {data/pseudo-reach-safe-si_sem.txt};
			\end{axis}
		\end{tikzpicture}
	}
	\subfloat[Parity]{
		\begin{tikzpicture}
			\begin{axis}[xmin=0,xmax=2100,ymax=1.0,samples=50,width=0.39\linewidth,height=5cm,
					axis x line = middle, axis y line* = middle,
					enlarge y limits=0,
					xlabel=\empty,ylabel=\empty,
					xtick={0,500,1000,1500,2000},xticklabels={,500,,1500},
					ytick={0,0.2,0.4,0.6,0.8,1.0},yticklabels={},
					legend pos={outer north east},
					cycle list name=no marks,
				]
				\addplot+ plot table[col sep=comma] {data/random-ltl-ql.txt};
				\addplot+ plot table[col sep=comma] {data/random-ltl-ql_prio.txt};
				\addplot+ plot table[col sep=comma] {data/random-ltl-ql_sem.txt};
				\addplot+ plot table[col sep=comma] {data/random-ltl-si.txt};
				\addplot+ plot table[col sep=comma] {data/random-ltl-si_sem.txt};
			\end{axis}
		\end{tikzpicture}
	}
	\caption{
		Detailed analysis of all considered methods on randomly generated games.
		We show the percentage of games solved within the given number of steps.
	} \label{fig:step_graph}
\end{figure}
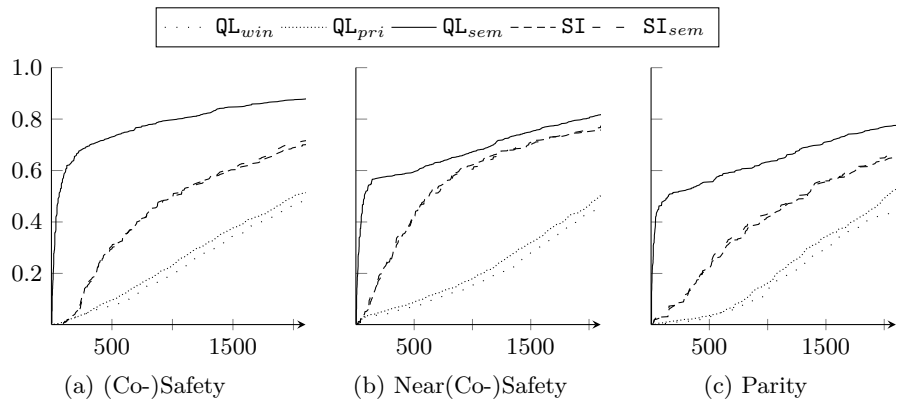

In \cref{tbl:results_simple}, we present our analysis of the trueness initialization on random formulae.
In particular, we compare $\texttt{SI}$ and $\texttt{SI}_{sem}$ with respect to how many games are solved by the initial strategy and the average solution size after the algorithm has converged.
In order to evaluate the initialization in $\texttt{SI}_{sem}$, we additionally break ties using the monitor labelling, taking the edge with the largest $\mathsf{progress}$ among those optimal w.r.t.\ the trueness of the master formula.
Our presented semantic initialization heuristic immediately identifies a winning strategy for more than half of the cases, while a randomly chosen strategy is only winning with roughly $10\%$ probability.
In particular, even in the more complex case of parity games, the trueness of the master formula proves to be a very good initialization heuristic.
The solution sizes do not differ significantly on these models, since the solutions usually are rather simple for such randomly generated formulas.
We highlight that for random initialization, there only is a negligible difference between the \enquote{Near(Co-)Safety} and \enquote{Parity} dataset, while our semantic approach works significantly better on the former set.

\begin{remark} \label{rem:onthefly}
	The performance of our initialization heuristic suggests interesting applications.
	For example, one could use this new initialization heuristic to explore and solve extremely large games.
	Observe that the game can be generated on the fly.
	Hence, when we initially follow trueness-optimal edges, we may immediately identify a solution while only constructing a small fraction of the game.
\end{remark}

\cref{fig:step_graph} shows an evaluation of all our methods on the randomly generated formulae in terms of evaluation steps.
The use of priorities consistently improves the performance of $\QL$, but only by a small amount.
On the other hand, the use of semantic information vastly improves the performance of $\QL$, outperforming even strategy improvement quite significantly.
Moreover, the difference between $\texttt{SI}$ and $\texttt{SI}_{sem}$ is negligible.
This is due to strategy improvement running until \emph{global} convergence, hence the algorithm spends effort in some unsolved regions of the game, even if the current strategy is already deciding optimally in most vertices.

\begin{table}[t]
	\setlength{\tabcolsep}{5pt}
	\centering
	\caption{
		Summary of our evaluation on the real-world data set of SYNTCOMP and several sets of randomly generated formulae.
		We show the geometric average of the number of evaluation steps on the respective data set.
		\enquote{Unrealizable} are all models in the SYNTCOMP data set which are not realizable.
		Out of the $\QL$ variants, we only include the average solution size for $\QL_{sem}$, since the solution sizes of all our $\QL$ methods are essentially equal.
	} \label{tbl:results_parity}
	\begin{tabular}{r|ccc|cc||ccc}
		                        &                    \multicolumn{5}{c||}{Avg. Eval. Steps}                     &      \multicolumn{3}{c}{Avg. Solution Size}       \\
		       Class (\#models) & $\QL_{win}$ & $\QL_{pri}$ & $\QL_{sem}$ & $\texttt{SI}$ & $\texttt{SI}_{sem}$ & $\QL_{sem}$ & $\texttt{SI}$ & $\texttt{SI}_{sem}$ \\
		\hline
		     \texttt{amba} (13) &    11540    &    9765     &    1271     &     1119      &        1089         &    78\%     &     73\%      &        46\%         \\
		     \texttt{lily} (23) &    2179     &    2052     &     168     &      639      &         580         &    21\%     &     21\%      &        27\%         \\
		  \texttt{ltl2dpa} (22) &    4909     &    3490     &    3944     &      561      &         552         &    44\%     &     36\%      &        18\%         \\
		\hline
		      Unrealizable (53) &    1141     &    1223     &     101     &      951      &         762         &     3\%     &      4\%      &         5\%         \\
		          Overall (206) &    3142     &    2664     &    1004     &      631      &         531         &    22\%     &     26\%      &        14\%         \\
		\hline\hline
		            (Co-)Safety &    2177     &    1993     &     103     &     1094      &        1060         &     7\%     &      7\%      &         7\%         \\
		        Near(Co-)Safety &    2243     &    1922     &     151     &      682      &         673         &    12\%     &     13\%      &        12\%         \\
		                 Parity &    2869     &    2141     &     174     &     1294      &        1157         &     7\%     &      8\%      &         9\%
	\end{tabular}
\end{table}

Inspired by these results on random games, we applied our algorithms to real-world problems.
The results are summarized in \cref{tbl:results_parity}.
In our experiments, we considered a large part of the SYNTCOMP set.
We additionally hand-picked some classes of formulae to discuss several observations about the semantic rewards.

The naive $\QL_{win}$ method is severely underperforming compared to other methods, as expected.
Moreover, $\QL_{sem}$ often outperforms both other $\QL$ variants.

The solution identified by the Q-learning methods often is larger than the one found by $\texttt{SI}_{sem}$.
We conjecture that this is due to Q-learning's bias towards exploration -- we did not incentivise the learner to yield small solutions.
The solution size practically is constant between the different $\QL$ methods, suggesting that these larger solutions are due to Q-learning itself.
Nevertheless, the solution size is comparable to the one of $\texttt{SI}$.
Moreover, we highlight that $\texttt{SI}_{sem}$'s solutions are significantly smaller than the one identified by $\texttt{SI}$, although the number of steps until convergence is essentially equal.
This suggests that our trueness initialization indeed identifies good initial strategies for such real-world games.

Another interesting observation is that our semantic approaches perform significantly better on unrealizable formulae, although these two cases theoretically are dual to each other.
We strongly conjecture that this is due to a bias in the data set.
Usually, unrealizable formulae are obtained by injecting small, local faults into an otherwise realizable formula.
These local faults are easy to find for our trueness / Q-learning approach.
This conjecture is strongly supported by the extremely small solutions found by all approaches.

The \texttt{lily} class seems to be of a similar structure, exhibiting fast convergence rates and small solutions.
On the \texttt{ltl2dpa} class, our semantic approach performs rather poorly compared to the priority based variant.
This class comprises unusually intricate temporal patterns. 
We conjecture that a more fine-tuned reward signal may improve performance especially on such models.


%% file: 8_conlusion.tex
\section{Conclusion}

We have presented the first step towards exploiting semantic labelling in LTL synthesis via the concept of trueness and the subsequent Q-learning.
By interpreting the labelling provided by semantic translations, the Q-learning agent can plan ahead instead of only seeing the next vertex.
Our first experimental evaluation already shows the potential of this idea.

\paragraph{Future work} includes several points of optimization.
Firstly, we want to provide a performant implementation of the on-the-fly exploration.
Once this is done, an in-depth performance comparison to state-of-the-art tools like \texttt{Strix} is desirable.
Furthermore, we can employ the semantic information to guide the exploration within these on-the-fly tools, as discussed in \cref{rem:onthefly}. 


Another interesting point is a more sophisticated definition of trueness.
Recall that our concept of trueness does not consider temporal aspects of the formula, yet it underpins all of our labelling-based approaches.
A different heuristic could yield significant improvements here.
Furthermore, one could use more complex learning methods instead of Q-learning.
These methods may among other things be able to re-use experience gained while solving a single game.


Finally, we plan on combining our learning methods with strategy iteration.
For example, the Q-learning agent can derive a good, but potentially not optimal strategy, and strategy iteration then solves the game with a few adjustments.

%

%% file: 9_appendix.tex
\section{Further data} \label{sec:appendix:data}

\begin{table}[htp]
	\setlength{\tabcolsep}{5pt}
	\caption{Priorities used to generate the random formulae.
		All unmentioned priorities are set to $0$.} \label{tbl:priorities}
	\centering
	\begin{tabular}{r|cccccc}
		           Class & $\land$ & $\lor$ & $\ltlGlobally$ & $\ltlFinally$ & $\ltlNext$ & $\ltlUntil$ \\
		\hline
		          Safety &    7    &   7    &       10       &       0       &     5      &      0      \\
		       Co-Safety &    7    &   7    &       0        &      10       &     5      &      0      \\
		   Pseudo-Safety &    7    &   7    &       10       &       1       &     5      &      1      \\
		Pseudo-Co-Safety &    7    &   7    &       1        &      10       &     5      &      1
	\end{tabular}
\end{table}

\begin{table}[htp]
	\setlength{\tabcolsep}{5pt}
	\caption{
		Overview of all timeouts for each considered algorithm and input class.
	}
	\centering
	\begin{tabular}{r|rrrrr}
		           Class & $\QL_{win}$ & $\QL_{pri}$ & $\QL_{sem}$ & $\texttt{SI}$ & $\texttt{SI}_{sem}$ \\
		\hline
		   \texttt{amba} &         8\% &         0\% &         0\% &           0\% &                 0\% \\
		   \texttt{lily} &         0\% &         0\% &         0\% &           0\% &                 0\% \\
		\texttt{ltl2dpa} &        11\% &         8\% &         7\% &           1\% &                 2\% \\
		\hline
		    Unrealizable &         4\% &         3\% &         2\% &          12\% &                25\% \\
		         Overall &        11\% &         7\% &         9\% &           4\% &                10\% \\
		\hline\hline
		     (Co-)Safety &         0\% &         0\% &         0\% &           0\% &                 0\% \\
		 Near(Co-)Safety &         1\% &         1\% &         1\% &           0\% &                 0\% \\
		          Parity &         1\% &         4\% &         3\% &           1\% &                 1\%
	\end{tabular}
\end{table}

\begin{figure}[htp]
	\centering
	\begin{tikzpicture}
		\begin{axis}[width=\textwidth,height=2cm,
				xlabel=\empty,ylabel=\empty,
				legend pos={outer north east},
				cycle list name=no marks,
				hide axis,
				xmin=0,xmax=0,ymin=0,ymax=0,
				legend style={
					at={(current bounding box.center)},
					anchor=center,
					legend columns=-1
				},
				legend entries={
					{$\QL_{win}$},
					{$\QL_{pri}$},
					{$\QL_{sem}$},
					{$\texttt{SI}$},
					{$\texttt{SI}_{sem}$},
				},
				cycle list name=no marks,
			]
			\pgfplotsinvokeforeach{1,...,5}{\addplot coordinates {(0,0)};}
		\end{axis}
	\end{tikzpicture}
	
	\subfloat[Safety]{
		\begin{tikzpicture}
			\begin{axis}[xmin=0,xmax=5000,ymax=1.0,samples=50,width=0.5\linewidth,height=5cm,
					axis x line = middle, axis y line* = middle,
					enlarge y limits=0,
					xlabel=\empty,ylabel=\empty,
					xtick={0,1000,2000,3000,4000},xticklabels={,1000,2000,3000,4000},
					ytick={0,0.2,0.4,0.6,0.8,1.0},yticklabels={},
					legend pos={outer north east},
					cycle list name=no marks
				]
				\addplot+ plot table[col sep=comma] {data/random-safe-ql.txt};
				\addplot+ plot table[col sep=comma] {data/random-safe-ql_prio.txt};
				\addplot+ plot table[col sep=comma] {data/random-safe-ql_sem.txt};
				\addplot+ plot table[col sep=comma] {data/random-safe-si.txt};
				\addplot+ plot table[col sep=comma] {data/random-safe-si_sem.txt};
			\end{axis}
		\end{tikzpicture}
	}
	\subfloat[Co-Safety]{
		\begin{tikzpicture}
			\begin{axis}[xmin=0,xmax=5000,ymax=1.0,samples=50,width=0.5\linewidth,height=5cm,
					axis x line = middle, axis y line* = middle,
					enlarge y limits=0,
					xlabel=\empty,ylabel=\empty,
					xtick={0,1000,2000,3000,4000},xticklabels={,1000,2000,3000,4000},
					ytick={0,0.2,0.4,0.6,0.8,1.0},yticklabels={},
					legend pos={outer north east},
					cycle list name=no marks
				]
				\addplot+ plot table[col sep=comma] {data/random-reach-ql.txt};
				\addplot+ plot table[col sep=comma] {data/random-reach-ql_prio.txt};
				\addplot+ plot table[col sep=comma] {data/random-reach-ql_sem.txt};
				\addplot+ plot table[col sep=comma] {data/random-reach-si.txt};
				\addplot+ plot table[col sep=comma] {data/random-reach-si_sem.txt};
			\end{axis}
		\end{tikzpicture}
	}

	\subfloat[Near Safety]{
		\begin{tikzpicture}
			\begin{axis}[xmin=0,xmax=5000,ymax=1.0,samples=50,width=0.5\linewidth,height=5cm,
					axis x line = middle, axis y line* = middle,
					enlarge y limits=0,
					xlabel=\empty,ylabel=\empty,
					xtick={0,1000,2000,3000,4000},xticklabels={,1000,2000,3000,4000},
					ytick={0,0.2,0.4,0.6,0.8,1.0},yticklabels={},
					legend pos={outer north east},
					cycle list name=no marks,
				]
				\addplot+ plot table[col sep=comma] {data/pseudo-safe-ql.txt};
				\addplot+ plot table[col sep=comma] {data/pseudo-safe-ql_prio.txt};
				\addplot+ plot table[col sep=comma] {data/pseudo-safe-ql_sem.txt};
				\addplot+ plot table[col sep=comma] {data/pseudo-safe-si.txt};
				\addplot+ plot table[col sep=comma] {data/pseudo-safe-si_sem.txt};
			\end{axis}
		\end{tikzpicture}
	}
	\subfloat[Near Co-Safety]{
		\begin{tikzpicture}
			\begin{axis}[xmin=0,xmax=5000,ymax=1.0,samples=50,width=0.5\linewidth,height=5cm,
					axis x line = middle, axis y line* = middle,
					enlarge y limits=0,
					xlabel=\empty,ylabel=\empty,
					xtick={0,1000,2000,3000,4000},xticklabels={,1000,2000,3000,4000},
					ytick={0,0.2,0.4,0.6,0.8,1.0},yticklabels={},
					legend pos={outer north east},
					cycle list name=no marks,
				]
				\addplot+ plot table[col sep=comma] {data/pseudo-reach-ql.txt};
				\addplot+ plot table[col sep=comma] {data/pseudo-reach-ql_prio.txt};
				\addplot+ plot table[col sep=comma] {data/pseudo-reach-ql_sem.txt};
				\addplot+ plot table[col sep=comma] {data/pseudo-reach-si.txt};
				\addplot+ plot table[col sep=comma] {data/pseudo-reach-si_sem.txt};
			\end{axis}
		\end{tikzpicture}
	}

	\caption{
		Data for the \enquote{(Co-)Safety} and \enquote{Near(Co-)Safety} sub-classes with the same notation as in \cref{fig:step_graph}.
	} \label{fig:step_graph_parts}
\end{figure}
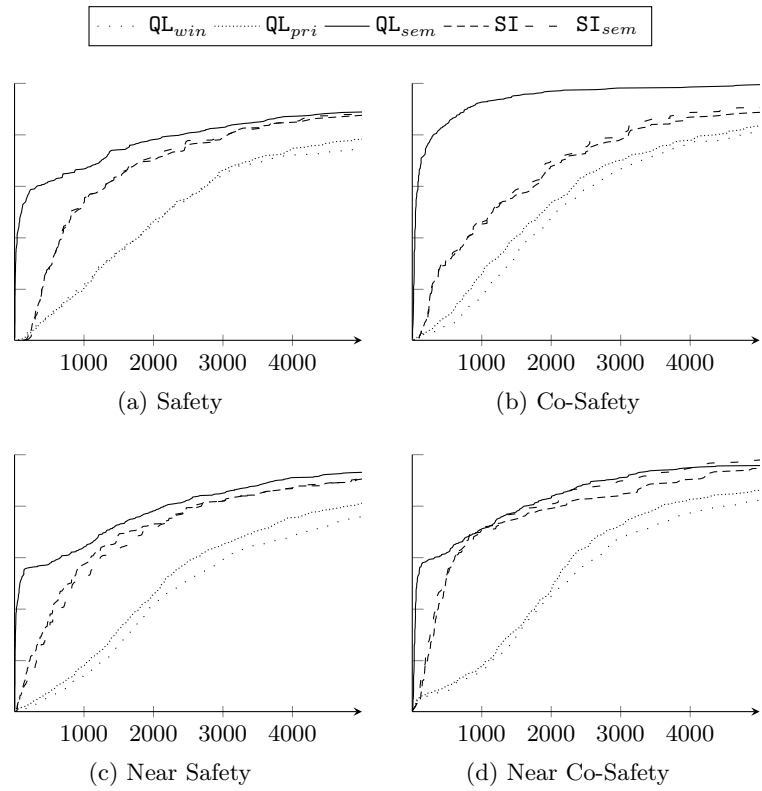

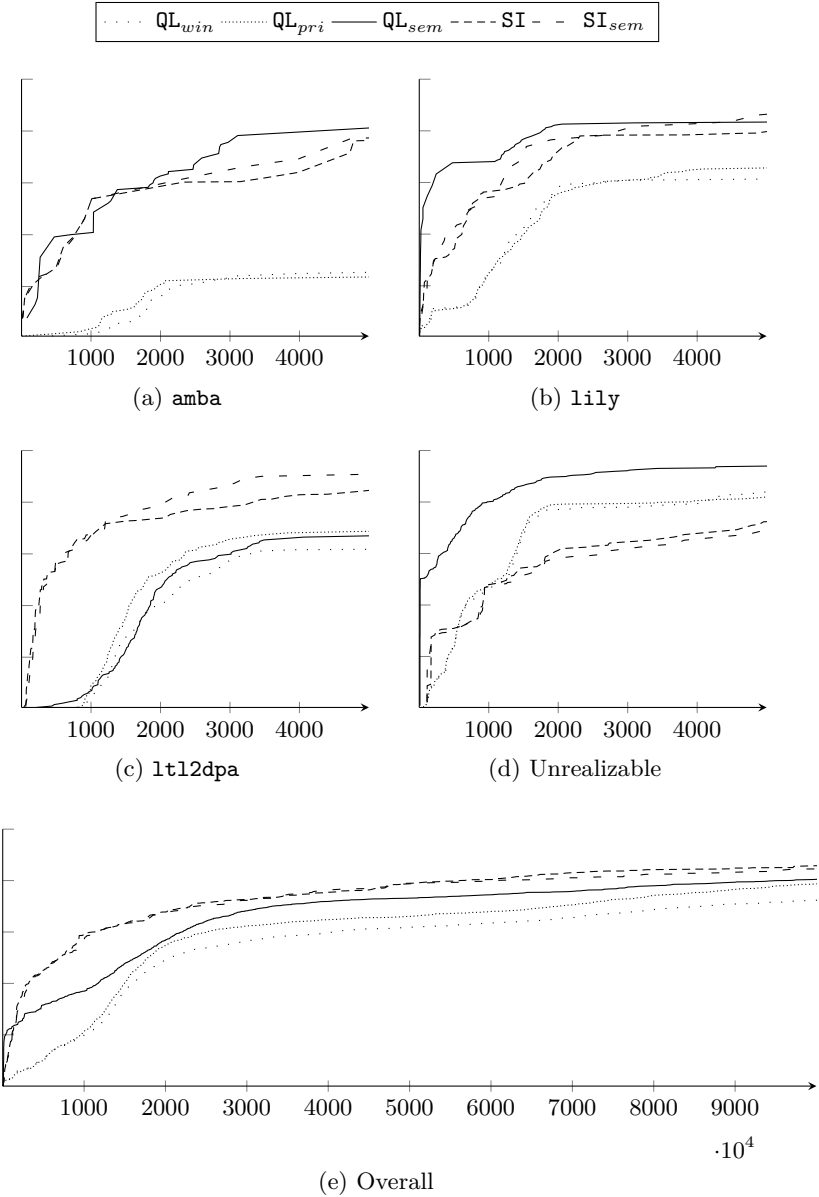
\begin{figure}[htp]
	\centering
	\begin{tikzpicture}
		\begin{axis}[width=\textwidth,height=2cm,
				xlabel=\empty,ylabel=\empty,
				legend pos={outer north east},
				cycle list name=no marks,
				hide axis,
				xmin=0,xmax=0,ymin=0,ymax=0,
				legend style={
					at={(current bounding box.center)},
					anchor=center,
					legend columns=-1
				},
				legend entries={
					{$\QL_{win}$},
					{$\QL_{pri}$},
					{$\QL_{sem}$},
					{$\texttt{SI}$},
					{$\texttt{SI}_{sem}$},
				},
				cycle list name=no marks,
			]
			\pgfplotsinvokeforeach{1,...,5}{\addplot coordinates {(0,0)};}
		\end{axis}
	\end{tikzpicture}
	
	\subfloat[\texttt{amba}]{
		\begin{tikzpicture}
			\begin{axis}[xmin=0,xmax=5000,ymax=1.0,samples=50,width=0.5\linewidth,height=5cm,
					axis x line = middle, axis y line* = middle,
					enlarge y limits=0,
					xlabel=\empty,ylabel=\empty,
					xtick={0,1000,2000,3000,4000},xticklabels={,1000,2000,3000,4000},
					ytick={0,0.2,0.4,0.6,0.8,1.0},yticklabels={},
					legend pos={outer north east},
					cycle list name=no marks
				]
				\addplot+ plot table[col sep=comma] {data/syntcomp-amba-ql.txt};
				\addplot+ plot table[col sep=comma] {data/syntcomp-amba-ql_prio.txt};
				\addplot+ plot table[col sep=comma] {data/syntcomp-amba-ql_sem.txt};
				\addplot+ plot table[col sep=comma] {data/syntcomp-amba-si.txt};
				\addplot+ plot table[col sep=comma] {data/syntcomp-amba-si_sem.txt};
			\end{axis}
		\end{tikzpicture}
	}
	\subfloat[\texttt{lily}]{
		\begin{tikzpicture}
			\begin{axis}[xmin=0,xmax=5000,ymax=1.0,samples=50,width=0.5\linewidth,height=5cm,
					axis x line = middle, axis y line* = middle,
					enlarge y limits=0,
					xlabel=\empty,ylabel=\empty,
					xtick={0,1000,2000,3000,4000},xticklabels={,1000,2000,3000,4000},
					ytick={0,0.2,0.4,0.6,0.8,1.0},yticklabels={},
					legend pos={outer north east},
					cycle list name=no marks
				]
				\addplot+ plot table[col sep=comma] {data/syntcomp-lily-ql.txt};
				\addplot+ plot table[col sep=comma] {data/syntcomp-lily-ql_prio.txt};
				\addplot+ plot table[col sep=comma] {data/syntcomp-lily-ql_sem.txt};
				\addplot+ plot table[col sep=comma] {data/syntcomp-lily-si.txt};
				\addplot+ plot table[col sep=comma] {data/syntcomp-lily-si_sem.txt};
			\end{axis}
		\end{tikzpicture}
	}

	\subfloat[\texttt{ltl2dpa}]{
		\begin{tikzpicture}
			\begin{axis}[xmin=0,xmax=5000,ymax=1.0,samples=50,width=0.5\linewidth,height=5cm,
					axis x line = middle, axis y line* = middle,
					enlarge y limits=0,
					xlabel=\empty,ylabel=\empty,
					xtick={0,1000,2000,3000,4000},xticklabels={,1000,2000,3000,4000},
					ytick={0,0.2,0.4,0.6,0.8,1.0},yticklabels={},
					legend pos={outer north east},
					cycle list name=no marks,
				]
				\addplot+ plot table[col sep=comma] {data/syntcomp-ltl2dpa-ql.txt};
				\addplot+ plot table[col sep=comma] {data/syntcomp-ltl2dpa-ql_prio.txt};
				\addplot+ plot table[col sep=comma] {data/syntcomp-ltl2dpa-ql_sem.txt};
				\addplot+ plot table[col sep=comma] {data/syntcomp-ltl2dpa-si.txt};
				\addplot+ plot table[col sep=comma] {data/syntcomp-ltl2dpa-si_sem.txt};
			\end{axis}
		\end{tikzpicture}
	}
	\subfloat[Unrealizable]{
		\begin{tikzpicture}
			\begin{axis}[xmin=0,xmax=5000,ymax=1.0,samples=50,width=0.5\linewidth,height=5cm,
					axis x line = middle, axis y line* = middle,
					enlarge y limits=0,
					xlabel=\empty,ylabel=\empty,
					xtick={0,1000,2000,3000,4000},xticklabels={,1000,2000,3000,4000},
					ytick={0,0.2,0.4,0.6,0.8,1.0},yticklabels={},
					legend pos={outer north east},
					cycle list name=no marks,
				]
				\addplot+ plot table[col sep=comma] {data/syntcomp-unrealizable-ql.txt};
				\addplot+ plot table[col sep=comma] {data/syntcomp-unrealizable-ql_prio.txt};
				\addplot+ plot table[col sep=comma] {data/syntcomp-unrealizable-ql_sem.txt};
				\addplot+ plot table[col sep=comma] {data/syntcomp-unrealizable-si.txt};
				\addplot+ plot table[col sep=comma] {data/syntcomp-unrealizable-si_sem.txt};
			\end{axis}
		\end{tikzpicture}
	}

	\subfloat[Overall]{
		\begin{tikzpicture}
			\begin{axis}[xmin=0,xmax=10000,ymax=1.0,samples=50,width=\linewidth,height=5cm,
					axis x line = middle, axis y line* = middle,
					enlarge y limits=0,
					xlabel=\empty,ylabel=\empty,
					xtick={0,1000,2000,3000,4000,5000,6000,7000,8000,9000},xticklabels={,1000,2000,3000,4000,5000,6000,7000,8000,9000},
					ytick={0,0.2,0.4,0.6,0.8,1.0},yticklabels={,0.2,0.4,0.6,0.8,1.0},
					legend pos={outer north east},
					cycle list name=no marks,
				]
				\addplot+ plot table[col sep=comma] {data/syntcomp-global-ql.txt};
				\addplot+ plot table[col sep=comma] {data/syntcomp-global-ql_prio.txt};
				\addplot+ plot table[col sep=comma] {data/syntcomp-global-ql_sem.txt};
				\addplot+ plot table[col sep=comma] {data/syntcomp-global-si.txt};
				\addplot+ plot table[col sep=comma] {data/syntcomp-global-si_sem.txt};
			\end{axis}
		\end{tikzpicture}
	}
	\caption{
		Data for all investigated SYNTCOMP classes with the same notation as in \cref{fig:step_graph}.
	} \label{fig:step_graph_synt}
\end{figure}

%% file: main.bbl
\begin{thebibliography}{10}

\bibitem{Syntcomp18}
The reactive synthesis competition: {SYNTCOMP} 2018 results.
\newblock http://www.syntcomp.org/syntcomp-2018-results/, 2018.

\bibitem{DBLP:conf/cav/BohyBFJR12}
Aaron Bohy, V{\'{e}}ronique Bruy{\`{e}}re, Emmanuel Filiot, Naiyong Jin, and
  Jean{-}Fran{\c{c}}ois Raskin.
\newblock Acacia+, a tool for {LTL} synthesis.
\newblock In {\em {CAV}}, 2012.

\bibitem{DBLP:journals/automatica/DingLB14}
Xu~Chu Ding, Mircea Lazar, and Calin Belta.
\newblock {LTL} receding horizon control for finite deterministic systems.
\newblock {\em Automatica}, 2014.

\bibitem{DBLP:conf/atva/Duret-LutzLFMRX16}
Alexandre Duret{-}Lutz, Alexandre Lewkowicz, Amaury Fauchille, Thibaud Michaud,
  Etienne Renault, and Laurent Xu.
\newblock Spot 2.0 - {A} framework for {LTL} and {\(\omega\)}-automata
  manipulation.
\newblock In {\em {ATVA}}, 2016.

\bibitem{DBLP:conf/cav/EsparzaK14}
Javier Esparza and Jan Kret{\'{\i}}nsk{\'{y}}.
\newblock From {LTL} to deterministic automata: {A} safraless compositional
  approach.
\newblock In {\em {CAV}}, 2014.

\bibitem{DBLP:conf/tacas/EsparzaKRS17}
Javier Esparza, Jan Kret{\'{\i}}nsk{\'{y}}, Jean{-}Fran{\c{c}}ois Raskin, and
  Salomon Sickert.
\newblock From {LTL} and limit-deterministic b{\"{u}}chi automata to
  deterministic parity automata.
\newblock In {\em {TACAS}}, 2017.

\bibitem{DBLP:conf/cav/FaymonvilleFT17}
Peter Faymonville, Bernd Finkbeiner, and Leander Tentrup.
\newblock Bosy: An experimentation framework for bounded synthesis.
\newblock In {\em {CAV}}, 2017.

\bibitem{DBLP:conf/cav/Fearnley17}
John Fearnley.
\newblock Efficient parallel strategy improvement for parity games.
\newblock In {\em {CAV}}, 2017.

\bibitem{DBLP:conf/atva/FriedmannL09}
Oliver Friedmann and Martin Lange.
\newblock Solving parity games in practice.
\newblock In {\em {ATVA}}, 2009.

\bibitem{DBLP:journals/corr/abs-1711-11439}
Swen Jacobs et~al.
\newblock The 4th reactive synthesis competition {(SYNTCOMP} 2017): Benchmarks,
  participants {\&} results.
\newblock In {\em SYNT@CAV}, 2017.

\bibitem{DBLP:conf/fmcad/JobstmannB06}
Barbara Jobstmann and Roderick Bloem.
\newblock Optimizations for {LTL} synthesis.
\newblock In {\em FMCAD}, 2006.

\bibitem{DBLP:conf/cav/JobstmannGWB07}
Barbara Jobstmann, Stefan~J. Galler, Martin Weiglhofer, and Roderick Bloem.
\newblock Anzu: A tool for property synthesis.
\newblock In {\em CAV}, 2007.

\bibitem{DBLP:conf/cav/KhalimovJB13}
Ayrat Khalimov, Swen Jacobs, and Roderick Bloem.
\newblock {PARTY} parameterized synthesis of token rings.
\newblock In {\em {CAV}}, 2013.

\bibitem{DBLP:journals/tcs/KleinB06}
Joachim Klein and Christel Baier.
\newblock Experiments with deterministic $\omega$-automata for formulas of
  linear temporal logic.
\newblock {\em Theor. Comput. Sci.}, 2006.

\bibitem{DBLP:conf/wia/KleinB07}
Joachim Klein and Christel Baier.
\newblock On-the-fly stuttering in the construction of deterministic
  $\omega$-automata.
\newblock In {\em CIAA}, 2007.

\bibitem{DBLP:conf/cav/KretinskyMSZ18}
Jan Kret{\'{\i}}nsk{\'{y}}, Tobias Meggendorfer, Salomon Sickert, and
  Christopher Ziegler.
\newblock Rabinizer 4: From {LTL} to your favourite deterministic automaton.
\newblock In {\em {CAV}}, 2018.

\bibitem{DBLP:conf/sofsem/Kupferman12}
Orna Kupferman.
\newblock Recent challenges and ideas in temporal synthesis.
\newblock In {\em SOFSEM}, 2012.

\bibitem{DBLP:conf/focs/KupfermanV05}
Orna Kupferman and Moshe~Y. Vardi.
\newblock Safraless decision procedures.
\newblock In {\em FOCS}, 2005.

\bibitem{DBLP:conf/cav/MeyerSL18}
Philipp~J. Meyer, Salomon Sickert, and Michael Luttenberger.
\newblock Strix: Explicit reactive synthesis strikes back!
\newblock In {\em {CAV}}, 2018.

\bibitem{michaud.18.synt}
Thibaud Michaud and Maximilien Colange.
\newblock Reactive synthesis from {LTL} specification with {S}pot.
\newblock In {\em Proceedings of the 7th Workshop on Synthesis, SYNT@CAV 2018},
  2018.

\bibitem{DBLP:conf/tacas/NeiderT16}
Daniel Neider and Ufuk Topcu.
\newblock An automaton learning approach to solving safety games over infinite
  graphs.
\newblock In {\em {TACAS}}, 2016.

\bibitem{DBLP:conf/lics/Piterman06}
Nir Piterman.
\newblock From nondeterministic {B}{\" u}chi and {S}treett automata to
  deterministic parity automata.
\newblock In {\em LICS}, 2006.

\bibitem{DBLP:conf/vmcai/PitermanPS06}
Nir Piterman, Amir Pnueli, and Yaniv Sa'ar.
\newblock Synthesis of reactive(1) designs.
\newblock In {\em VMCAI}, 2006.

\bibitem{DBLP:conf/focs/Safra88}
Shmuel Safra.
\newblock On the complexity of $\omega$-automata.
\newblock In {\em FOCS}, 1988.

\bibitem{DBLP:conf/fsttcs/Schewe07}
Sven Schewe.
\newblock Solving parity games in big steps.
\newblock In {\em {FSTTCS}}, 2007.

\bibitem{DBLP:conf/fossacs/Schewe09}
Sven Schewe.
\newblock Tighter bounds for the determinisation of {B}{\"u}chi automata.
\newblock In {\em FOSSACS}, 2009.

\bibitem{DBLP:conf/atva/ScheweF07a}
Sven Schewe and Bernd Finkbeiner.
\newblock Bounded synthesis.
\newblock In {\em {ATVA}}, 2007.

\bibitem{RL}
Richard~S. Sutton and Andrew~G. Barto.
\newblock {\em Reinforcement Learning: An Introduction}.
\newblock 2018.

\bibitem{DBLP:conf/tacas/Dijk18}
Tom van Dijk.
\newblock Oink: An implementation and evaluation of modern parity game solvers.
\newblock In {\em {TACAS}}, 2018.

\bibitem{DBLP:conf/lics/VardiW86}
Moshe~Y. Vardi and Pierre Wolper.
\newblock An automata-theoretic approach to automatic program verification
  (preliminary report).
\newblock In {\em {LICS}}, 1986.

\bibitem{DBLP:conf/cav/VogeJ00}
Jens V{\"{o}}ge and Marcin Jurdzinski.
\newblock A discrete strategy improvement algorithm for solving parity games.
\newblock In {\em {CAV}}, 2000.

\bibitem{DBLP:journals/tcs/Zielonka98}
Wieslaw Zielonka.
\newblock Infinite games on finitely coloured graphs with applications to
  automata on infinite trees.
\newblock {\em Theor. Comput. Sci.}, 1998.

\end{thebibliography}
